\newcommand{\EEFW}{\texttt{E2EFW}~}
\documentclass{webofc}
\usepackage[varg]{txfonts}   
\usepackage{xcolor}
\begin{document}
\title{End-to-end deep learning inference with CMSSW via ONNX using docker}
%
%

\author{\firstname{Purva} \lastname{Chaudhari}\inst{1} \and
        \firstname{Shravan} \lastname{Chaudhari}\inst{1} \and
        \firstname{Ruchi} \lastname{Chudasama}\inst{1} \thanks{\email{ruchi.chudasama@cern.ch}} \and
        \firstname{Sergei} \lastname{Gleyzer}\inst{1} on behalf of the CMS Collaboration
}

\institute{The University of Alabama }

\abstract{%
Deep learning techniques have been proven to provide excellent performance for a variety of high-energy physics applications, such as particle identification, event reconstruction and trigger operations. Recently, we developed an end-to-end deep learning approach to identify various particles using low-level detector information from high-energy collisions. These models will be incorporated in the CMS software framework (CMSSW) to enable their use for particle reconstruction or for trigger operation in real time. Incorporating these computational tools in the experimental framework presents new challenges. This paper reports an implementation of the end-to-end deep learning inference with the CMS software framework. The inference has been implemented on GPU for faster computation using ONNX. We have benchmarked the ONNX inference with GPU and CPU using NERSC’s Perlmutter cluster by building a docker image of the CMS software framework.
}
\maketitle
%
\section{Introduction}\label{intro}
The CMS~\cite{CMS} and ATLAS~\cite{ATLAS} experiments at the large hadron collider (LHC) have been designed to explore physics at the TeV energy scale during its operation span of about 30 years. Both experiments made a most significant discovery of the Higgs boson~\cite{ATLASHiggs,CMSHiggs} from the data collected at 8 TeV in 2012 worth integrated luminosity of 23 fb$^{-1}$.  In addition to the detailed studies of the Higgs boson properties, finding anomalies in the precision standard model (SM) to search for new physics with direct and indirect measurements are some of the important goals for the CMS experiment. The existence of the new particles as predicted by beyond standard model (BSM) theories are expected to have an extremely small production cross-section, necessitating the need to collect more collision data. The high-luminosity (HL-LHC)~\cite{HL-LHC} is planned to level the instantaneous luminosity at $5 \times 10^{34} \text{cm}^{-2} \text{s}^{-1}$, with an integrated luminosity of about 3000 fb$^{-1}$, ten times more than the LHC. The corresponding mean number of collisions (pileup) per bunch crossing will be 140 posing tremendous challenges for filtering, collecting, processing, reconstructing, and analyzing data due to huge event size, data volume, and complexity. 

Advanced machine learning approaches will be employed to overcome significant hurdles provided by rising levels of pile-up and the scarcity of sought-after signals to achieve the key physics goals at HL-LHC. To address this issue, CMS researchers are employing cutting-edge machine learning techniques for data processing and detector reconstruction, with the goal of optimizing and speeding up these models during training and inference. Most of the particle identification algorithms at the CMS and ATLAS experiments rely on inputs provided by the particle-flow (PF)~\cite{PF} algorithm used to convert detector-level information to physics objects due to its capability to significantly reduce the size and complexity of particle physics data while offering a physically intuitive and simple-to-use representation for physics analyses. Despite the very high reconstruction efficiency of PF algorithms, some physics objects may fail to be reconstructed, are reconstructed imperfectly, or exist as fakes and limit the search for BSM scenarios. Therefore it is advantageous to consider reconstruction that allows a direct application of machine learning algorithms to low-level data in the detector. 

The end-to-end deep learning approach combines a low-level detector representation and deep learning algorithms. This approach has achieved current state-of-the-art performance in identifying electrons, photons, jets, and boosted objects \cite{E2E:Egamma,E2E:qg,E2E:BoostedTop,E2E:GNN} using various deep-learning architecture such as convolutional neural network and graph neural network. One of the main objectives of the CMS experiment's research and development towards HL-LHC is to incorporate such cutting-edge machine learning algorithms for particle identification into the CMS software framework (CMSSW)~\cite{CMSSW} data processing pipeline. Training the deep neural networks and subsequently obtaining the inference on trained models on massive amounts of data, such as those produced at LHC, is exceedingly time-consuming and demands significant computational resources. Graphical Processing Units have proven to be capable of providing fast, parallelized, and energy-efficient processing of data even on complex deep-learning architectures, making them ideal for a wide range of real-time applications as well as for user analysis-specific tasks. 

This paper presents the integration of end-to-end deep learning framework into CMSSW to discriminate electrons from photons (\texttt{E/Gamma tagger}), quark jets from gluon jets (\texttt{Quark/Gluon tagger}), top quark from QCD jets (\texttt{Top tagger}), hadronic taus from QCD jets (\texttt{Tau tagger}). Their inference times are benchmarked on CPU and GPU.  


\section{Simulated dataset for end-to-end deep learning}
The end-to-end deep learning technique is based on high-fidelity Monte Carlo simulated event samples. The samples are produced for the 2018 proton-proton collisions data-taking periods at the center of mass energy 13 TeV, without considering any additional particle collision in the single bunch crossing. 
The events for \texttt{E/Gamma tagger} studies are generated with photon particle gun sample at transverse momentum, $p_{\rm T} = 50$ GeV. The multijet production of light-flavor and gluon jets via strong interaction, referred to as quantum chromodynamics (QCD) with hard scattered transverse momentum $\hat{p_{\rm T}}$ between 300 to 470 GeV are generated with PYTHIA 8~\cite{PYTHIA} for \texttt{Quark/Gluon tagger} studies. Events from top quark-antiquark pair production where W boson from the top quark decay are required to decay as a quark are used for \texttt{Top tagger}.  The monte carlo samples were generated with \textsc{POWHEG} v2.0~\cite{POWHEG} at next to leading order in perturbative QCD and uses \textsc{PYTHIA} 8 for Parton showering. The generation of 125 GeV Higgs boson (H) events via gluon fusion at NLO with H decays to tau leptons (H $\to\tau\tau$) is also performed with the POWHEG 2.0 generator for \texttt{Tau tagger} studies. 

The LHC provides countercirculating beams of high-energy protons, allowing bunches of protons in these beams to interact with each other in the CMS detector \cite{CMS} every 25 ns. When protons from opposing beams collide, a wide range of physical processes can occur, leading to the formation of either fundamental or composite particles. These particles, or their decay products, can then enter the CMS detector, which is designed to determine the particle type, energy, and momentum. Each bunch crossing where proton collisions occur is referred to as an "event," which can project hundreds of particles into the CMS detector. The CMS detector is designed as a series of concentric cylindrical sections with barrel and endcap sections that enclose a primary interaction point where the LHC proton beams collide with each other. The central feature of the CMS detector is a superconducting solenoid that provides a magnetic field of 3.8 T designed to bend the trajectories of charged particles that aid in transverse momentum, $p_{\rm T}$ measurement. The silicon tracker is composed of two parts, namely, the silicon pixel detector and the silicon strip detector. The first silicon pixel detector is the innermost part and is composed of four layers in the barrel region \textsc{(BPIX)} and three disks in the endcap region \textsc{(FPIX)}. The pixel detector provides crucial information for vertexing and track seeding. The outer part of the tracking system is composed of silicon strip detectors. This is followed by the electromagnetic calorimeter \textsc{(ECAL)}, made of lead-tungstate crystals, to measure the energy of  electromagnetically interacting particles, and then the hadronic calorimeter \textsc{(HCAL)}, made of brass towers, to measure the energy of hadrons. These are surrounded by the solenoid magnet which is finally encased by the muon chambers to detect the passage of muons.

The detector response for all samples was simulated for CMSSW release 12\_0\_2 using \textsc{GEANT 4} package, which delivers the state-of-the-art in ﬁrst-principles detector simulation, along with the most detailed geometry models of the CMS detector. For this study, we additionally use a custom CMS data format which includes the low-level tracker detector information, specifically, the reconstructed clusters from the pixel and silicon strip detectors. 

\section{Integration of end-to-end deep learning with CMS software framework}
CMSSW is an overall collection of software framework written in C++ that is built around the Event Data Model \textsc{(EDM)}. In the \textsc{(EDM)} format, each entry \texttt{(edm::Event)} represents a single collision event capable of holding multiple attributes on top of which various modules can be run to perform simulation, digitization, and reconstruction. These modules are divided according to their functionality into \texttt{CMSSW} base classes for analyzing data collections \texttt{(edm::EDAnalyzer)} or producing new ones \texttt{(edm::EDProducers)}, among many other modules. The end-to-end framework (\EEFW) is designed to be highly modular and adaptable in order to accommodate customized workflows for end-to-end ML training and inference. The \EEFW can be optionally run to produce EDM-format files for further downstream processing by other CMSSW modules for a production workflow or to produce ROOT-ntuples for rapid prototyping. 

\subsection{End-to-end framework pipeline}

\begin{figure}[!htbp]
\centering
\includegraphics[width=0.99\linewidth]{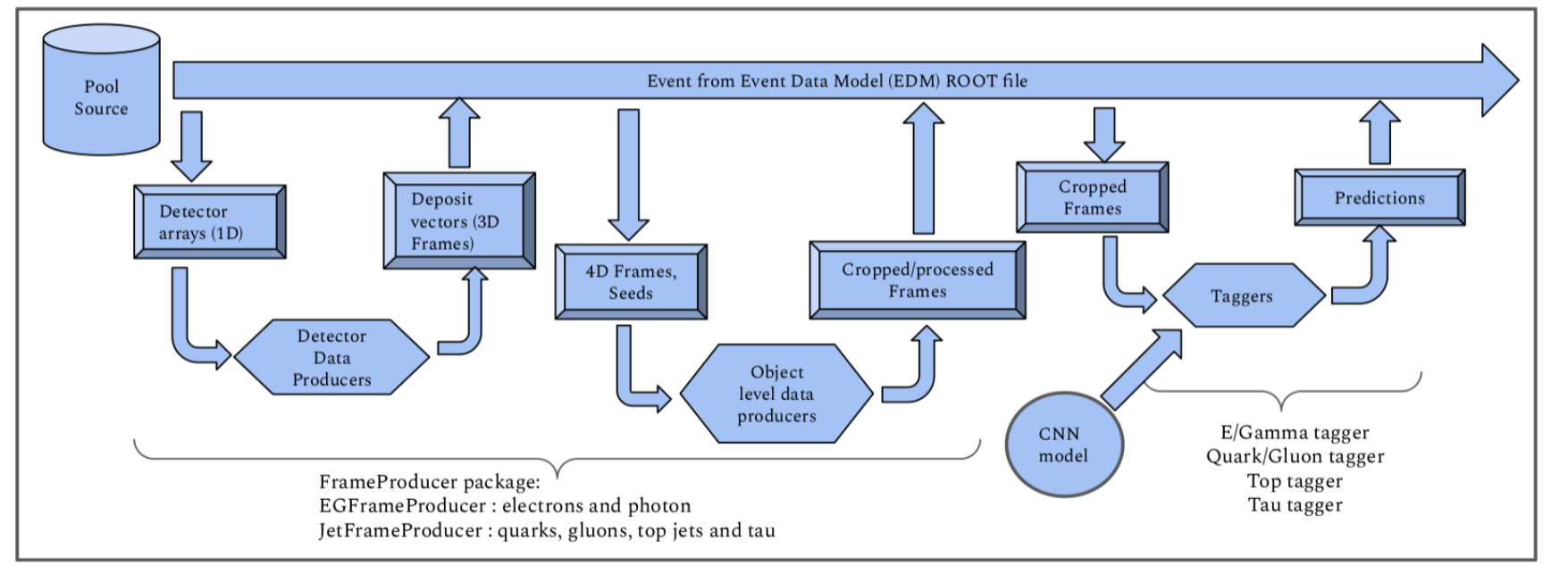}
\caption{ The end-to-end framework (\EEFW) pipeline used for \texttt{E/Gamma,Quark/Gluon,Top,Tau} taggers \cite{DPSE2E}.}
\label{fig:EEFW}
\end{figure} 

\EEFW consists of three main package categories: \texttt{DataFormats}, \texttt{FrameProducers,} and \texttt{Taggers}. The \texttt{DataFormats} package consists of all the objects and classes required to execute the \EEFW modules and save the output back to EDM-format files. It contains convenience classes for dealing with inputs and defining associations with other relevant collections. The association maps for linking object-level detector inputs to their related reconstructed physics objects are specifically defined here.

The \texttt{FrameProducer} package is primarily responsible for extracting detector data either as whole event data or as object-level data—and auxiliary functions aiding in this regard. The \textit{data producers} are provided with hit information from various CMS subdetectors in order to cast the whole event data to desired images or graphs. The \textit{object-level data producers} enable the generation of multi-subdetector data, such as to produce images or graphs for reconstructed electrons, photons, and jets as used in this paper. There are also a variety of modules that integrate with the output of \textit{detector data producers} to create localized windows or crops around the coordinates of a desired reconstructed physical object. The \EEFW has user-configurable options for controlling which subdetectors to include and which jet clustering technique to employ to determine their centers. A separate module is given for electrons and photons. As a result, depending on the user's needs, different combinations of the producers might be utilized for different tasks. For instance, \texttt{EGFrameProducer} is used to produce electron/photon showers for \texttt{EGTagger} while \texttt{JetFrameProducer} is used for \texttt{Quark/Gluon,Top and Tau tagger}. 

More detailed analyses of the reconstructed objects will be required for the typical physics application. To support this purpose, a third  package category \texttt{Taggers} is provided. These can be used to interface with the output of any preceding \texttt{FrameProducer} package, allowing for modular, highly customized workflows. While most production-level analyses will have their own dedicated analysis workflow, the \texttt{Taggers} provide a quick and convenient avenue for rapid prototyping and analysis, which is desirable during the ML algorithm development but can also be used for running inference in a production-like workflow. The \EEFW presented in this paper supports a number of template modules. 

 A typical end-to-end framework pipeline is illustrated in Figure \ref{fig:EEFW}. First, the detector-level data producer is run to extract whole detector images or graphs corresponding to the ECAL, HCAL, and Track layers and store these back into an EDM-format file in vector shape suitable for object-level cropping. Then, object-level data producers are run to crop and process these whole detector vectors into either photon-level or jet-level data around the coordinates of the reconstructed photons or jets and again push back these vectors into the EDM-format file. Photon-level objects are used only for \texttt{E/Gamma tagger} while jet-level objects are used for all other taggers described in this paper.  The Tagger package is configured to use various deep-learning architectures such as Convolutional Neural Network (CNN) (used in this study), and Vision Transformers for images. We considered simpleNet CNN architecture to benchmark the inference for all four taggers. The information from one sub-detector is considered as one channel of the CNN model. The input tensor size and the 
 number of channels can be configured according to the end-user necessity. The \EEFW provides an option to consider 13 channels, such as Track $p_{\rm T}$, d$_{0}$, d$_{z}$, four \textsc{BPIX} layers, \textsc{ECAL \& HCAL}, and four strip layers. Table~\ref{tab-1} shows various sub-detector channels and input tensors considered for the various Taggers.  The \EEFW can be easily configured to use other algorithms like Graph Neural Networks (GNNs) with graph inputs, which is beyond the scope of this paper.
 
 \begin{table}[htb]
\caption{Specifications of the convolutional neural network inputs.}
\label{tab-1}       
\centering
\begin{tabular}{l | l | l | l}
\hline
Tagger   & Number  & Input tensor & Channel names \\
         &  of channels & array size & \\\hline
\texttt{E/Gamma}  & 1   &  $1 \times 32 \times 32$ & \textsc{ECAL} \\
\texttt{Quark/Gluon}  & 5   &  $5 \times 128 \times 128$ & Track $p_{\rm T}$, d$_{0}$, d$_{z}$, \textsc{ECAL \& HCAL} \\
\texttt{Top}  & 8   &  $8 \times 128 \times 128$ & Track $p_{\rm T}$, d$_{0}$, d$_{z}$, \textsc{BPIX} layers, \textsc{ECAL \& HCAL} \\
\texttt{Tau}  & 8   &  $8 \times 128 \times 128$ & Track $p_{\rm T}$, d$_{0}$, d$_{z}$, \textsc{BPIX} layers, \textsc{ECAL \& HCAL} \\\hline
\end{tabular}
\end{table}

 Lastly either one of the four taggers is run to select either photon or reconstructed jet. The data associated with the selected photon-level or jet-level objects are passed to the \textsc{ONNX}~\cite{Onnex} Runtime for inference. The resulting predictions on these selected photons or jets are then pushed back into the EDM-format file for further downstream analysis. ONNX (Open Neural Network Exchange) is an open-source framework designed to facilitate interoperability and portability between different deep learning frameworks and tools. Most deep learning frameworks such as TensorFlow, PyTorch, XGBoost support converting those models into the ONNX format or loading a model from an ONNX format. \textsc{ONNX} Runtime is used for inference on \textsc{ONNX} models as it is supported by the CMSSW to run the inference on GPUs. 
While we have described photon-level and jet-level workflow, the \EEFW can be appropriately adapted to process other objects and full events, by a suitable definition of the process workflow. For instance, if a user wishes to perform event-level analysis and inference on whole detector inputs, an appropriate Tagger package that applies event-level selection can be defined that interfaces directly with the output of detector data producer and bypasses object-level data producers. The \EEFW thus allows a high degree of flexibility for pursuing an assortment of end-to-end ML studies dictated by the user. 

\subsection{Containerising end-to-end framework at NERSC using docker}
Despite the dedicated computing environment provided by CMS experiment, the installation of CMS software framework and use on a third-party computing facility is still very tedious, error-prone, and time-consuming, due to the versioning issues and lack of administrative privileges. To address this issue, our group has demonstrated the efficient use of container images and their application while performing analysis on the Perlmutter supercomputing facility at the National Energy Research Scientific Computing Center (NERSC)~\cite{Nersc}. Docker is one such well-known and versatile unit of software for effectively managing dependencies, runtime, system tools, system libraries, and settings. Virtualization and isolation of software packages provide a great deal of flexibility during the development stage, in reliably deploying and efficiently managing pipelines. We dockerized CMSSW version 12\_0\_6 and the required dependencies in a compact docker image that was reliably transferred to the Perlmutter cluster. Perlmutter cluster provides the Cern Virtual Machine File System (CVMFS) required to execute the CMS software framework. The shifter software package at NERSC allows to run the docker image of CMSSW or any other user-created images. 


\section{Computing resources}
The end-to-end framework inference has been benchmarked at two different computing sites. Fermilab LHC Physics Center (LPC) provides NVIDIA Tesla P100 GPU that utilizes the Pascal architecture. The Tesla P100 \cite{P100} GPU with high bandwidth memory (HBM) of 12 GB was accessed at Fermilab via a dedicated GPU worker node through a 12 GB/s PCIe connection using CUDA Toolkit driver version 12.1. Data was stored and read from an HGST 1W10002 hard drive located on the GPU machine during inference. Images produced from the end-to-end framework were provided to the GPU using a single Intel(R) Xeon(R) Silver 4110 8-core CPU. LPC also provides CPU-only resources, the \EEFW inference was benchmarked on AMD EPYC Processor 1-core CPU. We also used the GPU resources located at the Perlmutter cluster at the National Energy Research Scientific Computing Center (NERSC). Perlmutter cluster provides NVIDIA A100~\cite{A100} GPU that utilizes Ampere architecture with high bandwidth memory (HBM) of 40 GB and it was accessed through a dedicated GPU worker node through a 25 GB/s PCIe connection using CUDA Toolkit driver version 11.7. Data was stored and read from an Intel Dual-Port NVMe solid-state drive. 
 
\section{Performance}
\begin{figure}[!htbp]
\centering
\includegraphics[width=0.80 \linewidth]{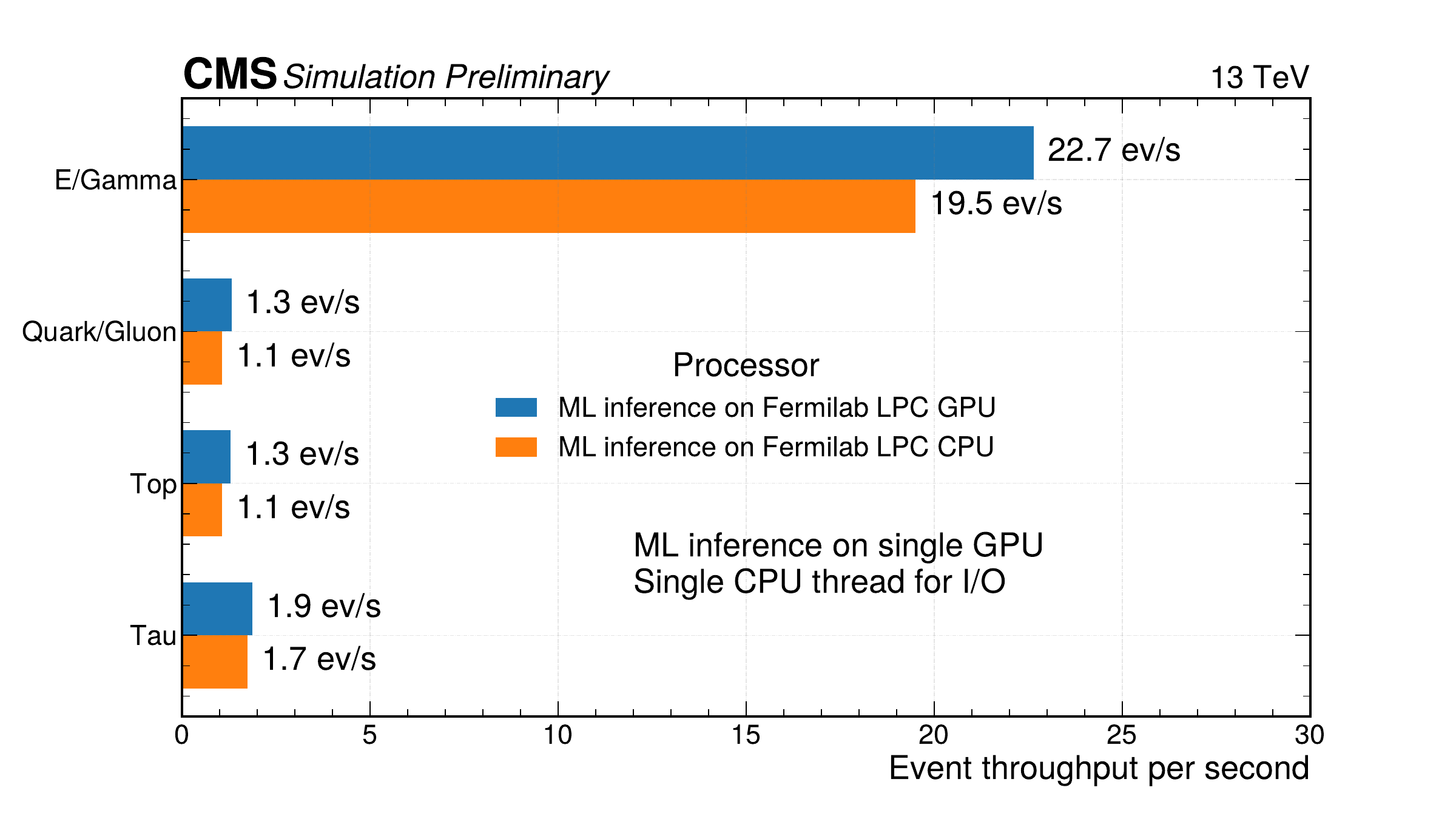}
\caption{ End-to-end inference framework event throughput per second for \texttt{E/Gamma, Quark/Gluon, Top, and Tau taggers} compared for Fermilab LPC GPU (blue) and CPU (orange) \cite{DPSE2E}.}
\label{fig:throughputCPU-GPU}
\end{figure}

\begin{figure}[!htbp]
\centering
\includegraphics[width=0.80 \linewidth]{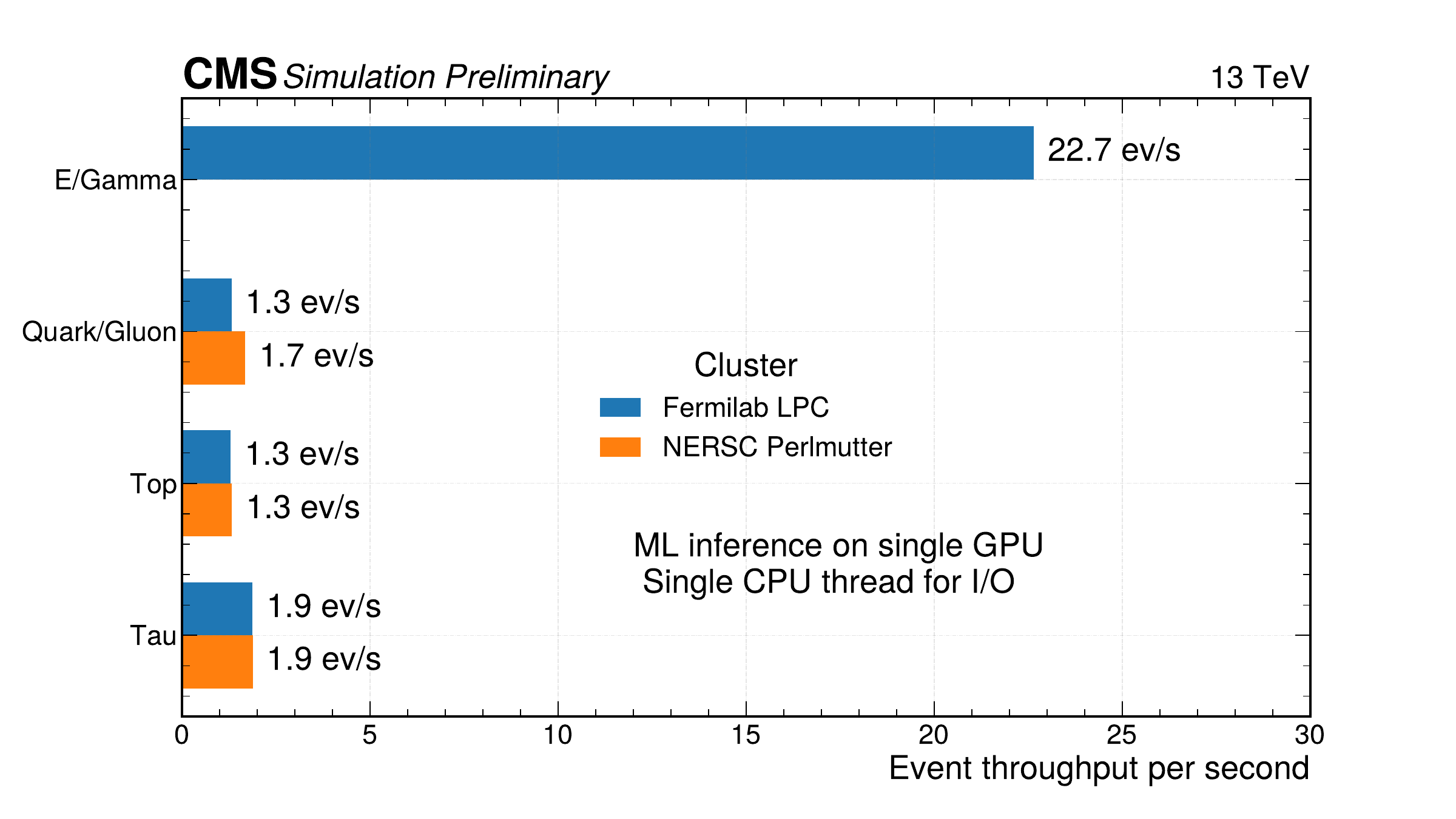}
\caption{End-to-end inference framework event throughput per second for \texttt{E/Gamma, Quark/Gluon, Top, and Tau taggers} compared for NVIDIA Tesla P100 GPU at Fermilab LPC (blue) and NVIDIA A100 GPU at NERSC Perlmutter (orange) \cite{DPSE2E}.}
\label{fig:throughputGPUPerl}
\end{figure}

The latency and throughput for \EEFW framework for inference were obtained by running 1000 events, generated from the Monte Carlo simulations as described in section 2. The inference was obtained by running a single-threaded job out of which the first 300 events were dropped from the calculations to stabilize the results. The measurement was repeated ten times and the average throughput value was estimated. Figure \ref{fig:throughputCPU-GPU} shows the average end-to-end inference framework event throughput per second for \texttt{E/Gamma,Quark/Gluon,Top,and Tau taggers} compared for Fermilab LPC GPU and CPU. The ML inference was obtained on a single GPU and single CPU with a single thread for input/output. The speedup of 11-18\% was achieved for \EEFW inference on Fermilab GPU compared to CPU achieved due to the 8-core CPU and an NVIDIA Tesla P100 GPU available on the LPC GPU node compared to the 1-core CPU on the LPC CPU. Figure \ref{fig:throughputGPUPerl} compares the average throughput per second for NVIDIA Tesla P100 GPU at Fermilab LPC and NVIDIA A100 GPU at NERSC for \texttt{E/Gamma, Quark/Gluon, Top, and Tau taggers}. The figure presently does not include the results for \texttt{E/Gamma tagger} for A100 GPU at NERSC Perlmutter. Figure \ref{fig:throughputGPUPerl} demonstrates a small increase for \texttt{Top} and \texttt{Tau Tagger} and a 20\% improvement for \texttt{Quark/Gluon tagger} when utilizing A100 GPU against P100. This is because the inference at Perlmutter is obtained by first executing the CMSSW docker image, which takes around 5\% of the entire inference time, and then opening an input file, which takes more than 50\%. Aside from that, the \texttt{Tagger} package's ONNX Runtime inference is the only component currently implemented on the GPU. Future plan is to port other modules of the framework to the GPU as well. These throughput measurements have 0.5-3\% uncertainties.

\begin{figure}[!htbp]
\centering
\includegraphics[width=0.49\linewidth]{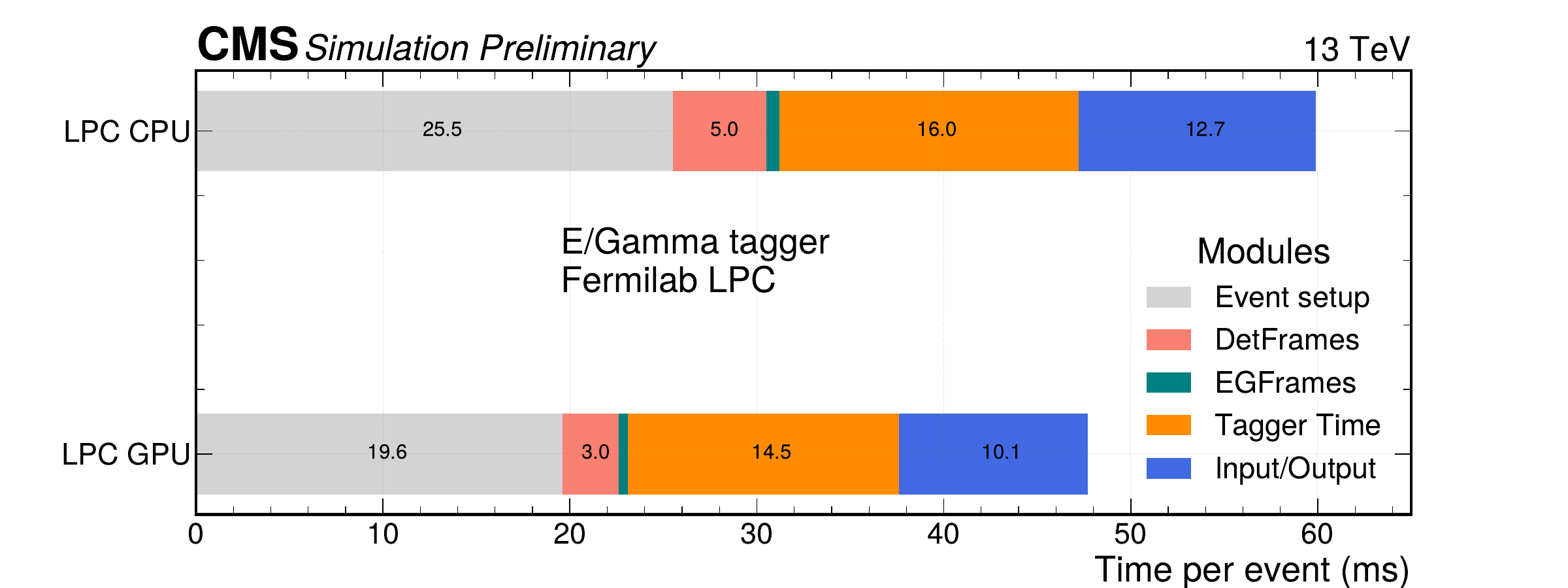}
\includegraphics[width=0.49\linewidth]{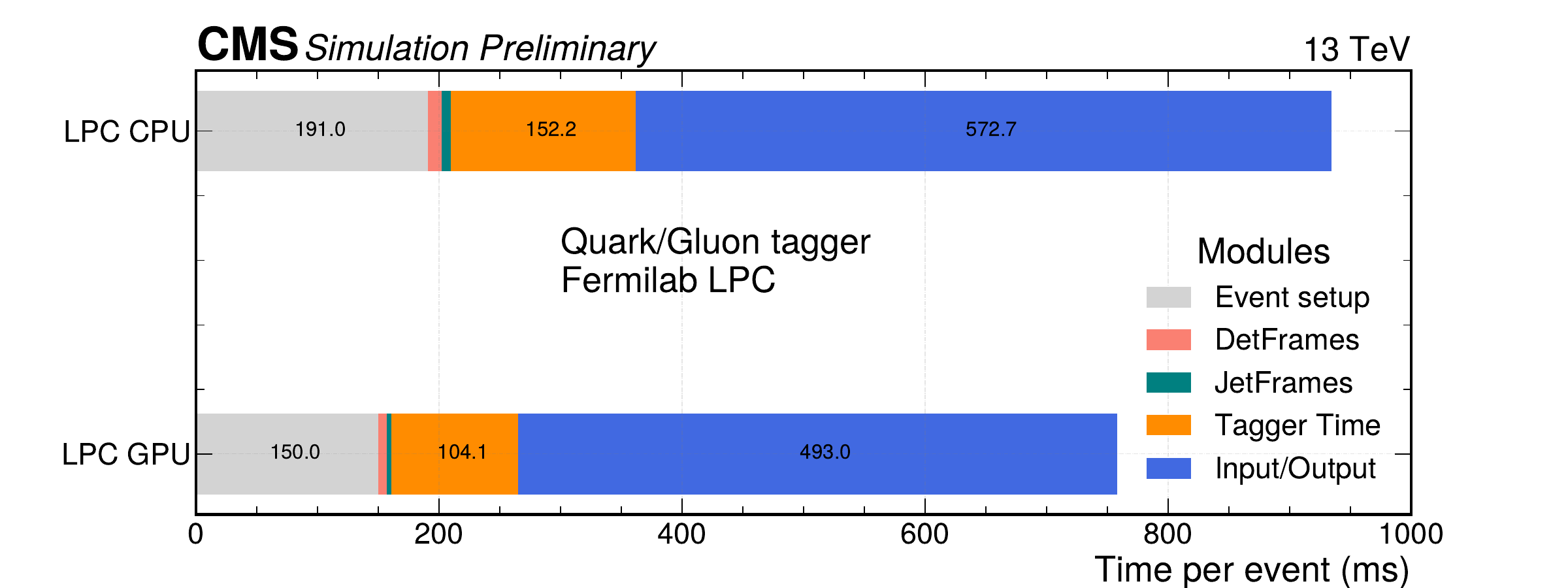}
\includegraphics[width=0.49\linewidth]{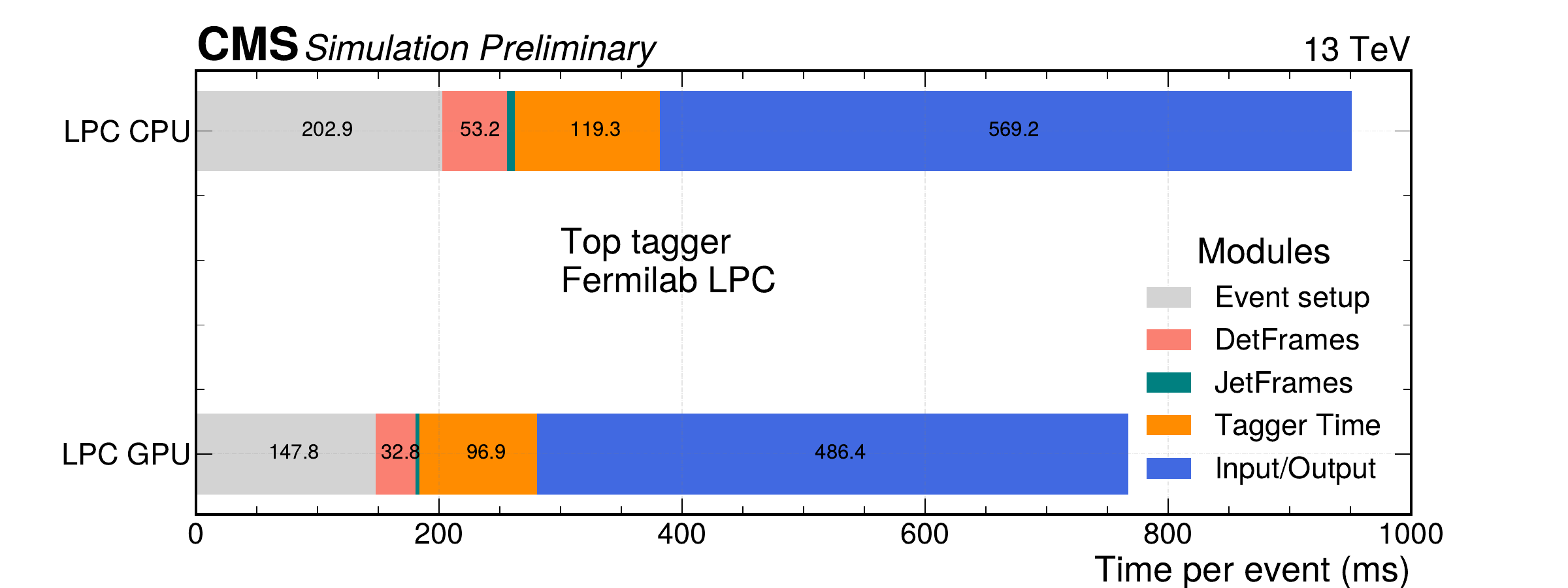}
\includegraphics[width=0.48\linewidth]{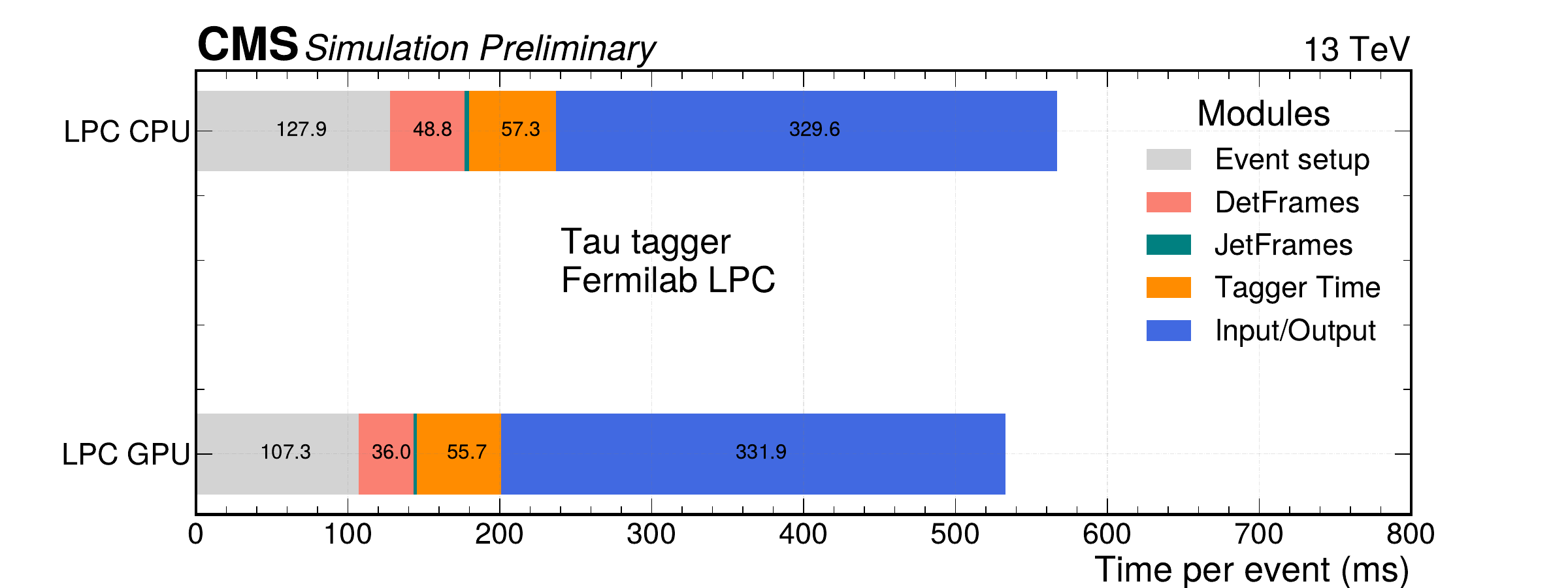}
\caption{ Time spent by end-to-end inference framework modules per event  such as Event setup (gray), DetFrames (Pink), EGFrames /JetFrames (Teal), Tagger time (orange), compared for Fermilab LPC CPU and GPU for \texttt{E/Gamma tagger} (top left), \texttt{Quark/Gluon tagger} (top right), \texttt{Top tagger} (bottom left), and \texttt{Tau tagger} (bottom right) \cite{DPSE2E}.}
\label{fig:egtiming}
\end{figure}

Figure ~\ref{fig:egtiming} compares the breakdown of time spent per event (latency) by the end-to-end inference framework modules, such as Event setup (gray), DetFrames (Pink), EGFrames /JetFrames (Teal), Tagger time (orange), and input/output (blue) at LPC CPU in top bar chart and LPC GPU in the bottom bar chart. The timings are compared for \texttt{E/Gamma,Quark/Gluon,Top and Tau Tagger}, respectively. Figure ~\ref{fig:egtiming}(top left) shows that around 40\% of the time per event was spent in event setup for \texttt{E/Gamma tagger} while ~60\% of time was spent in Input/Output module for {Quark/Gluon, Top and Tau} tagger.  The I/O time can be improved further by writing only the required information in the EDM-format root files instead of the full input information plus the data producers and object producer collections produced in the end-to-end framework.

\section{Conclusion}
This paper presents the integration of end-to-end deep learning framework within CMS software framework. The framework provides the support to discriminate electrons from photon showers, quark jets from gluon jets, top quark from QCD jets, and tau particles from QCD jets. Along with the particle identification task on photon-level or jet-level, the framework can be easily adapted to even-level tasks. We utilized the Fermilab LPC and at the NERSC Perlmutter's Central Processing Unit and Graphical Processing units to benchmark the inferences for \texttt{E/Gamma,Quark/Gluon,Top,and Tau taggers}. The inference obtained on NVIDIA Tesla P100 GPU with a single Intel(R) Xeon(R) Silver 4110 8-core CPU shows 11-16\% increase in throughput compared to AMD EPYC Processor 1-core CPU. The integration of the end-to-end deep learning framework within the CMS software framework was accomplished at the NERSC supercomputing facility by preparing a CMSSW docker image. Then the inference was benchmarked on NVIDIA Tesla A100 GPU, demonstrating a small increase for \texttt{Top} and \texttt{Tau} taggers and a 20\% improvement for \texttt{Quark/Gluon} tagger against NVIDIA Tesla P100 GPU. 

\section*{Acknowledgments}
We would like to thank the CMS Collaboration and the CERN. This work is supported in part by the U.S. Department of Energy (DOE) under award no. DE-SC0012447. This work is partially supported by the Fermilab US-CMS HL-LHC Software and Computing R\&D efforts. This research used resources from the National Energy Research Scientific Computing Center (NERSC), a U.S. Department of Energy Office of Science User Facility located at Lawrence Berkeley National Laboratory. PC and SC were participants in the Google Summer of Code program in 2021-2022 and 2020-2021, respectively.

%
%

%

\end{document}